\documentclass{sigchi}

\CopyrightYear{2017}
\setcopyright{acmlicensed}
\conferenceinfo{CSCW '17,}{February 25-March 01, 2017, Portland, OR, USA}
\isbn{978-1-4503-4335-0/17/03}
\doi{http://dx.doi.org/10.1145/2998181.2998345}

\usepackage{balance}  
\usepackage{graphics} 
\usepackage{txfonts}
\usepackage{times}    
\usepackage[pdftex]{hyperref}
\usepackage{color}
\usepackage{textcomp}
\usepackage{booktabs}
\usepackage{ccicons}
\usepackage{todonotes}
\usepackage[inline]{enumitem}

\makeatletter
\def\url@leostyle{%
  \@ifundefined{selectfont}{\def\UrlFont{\sf}}{\def\UrlFont{\small\bf\ttfamily}}}
\makeatother
\def\pprw{8.5in}
\def\pprh{11in}

\definecolor{linkColor}{RGB}{6,125,233}
\hypersetup{%
  pdftitle={Bots, Seeds and People: Web Archives as Infrastructure},
  pdfauthor={Ed Summers and Ricard Punzalan},
  pdfkeywords={archive; web; collaboration; design; practice},
  bookmarksnumbered,
  pdfstartview={FitH},
  colorlinks,
  citecolor=black,
  filecolor=black,
  linkcolor=black,
  urlcolor=linkColor,
  breaklinks=true,
}

\begin{document}

\title{Bots, Seeds and People}
\subtitle{Web Archives as Infrastructure}

\numberofauthors{3}
\author{
  \alignauthor Ed Summers \\
      \affaddr{University of Maryland}\\
       \email{edsu@umd.edu}
\and   \alignauthor Ricardo Punzalan \\
      \affaddr{University of Maryland}\\
       \email{punzalan@umd.edu}
\and }

\maketitle

\begin{abstract}
The field of web archiving provides a unique mix of human and automated
agents collaborating to achieve the preservation of the web. Centuries
old theories of archival appraisal are being transplanted into the
sociotechnical environment of the World Wide Web with varying degrees of
success. The work of the archivist and bots in contact with the material
of the web present a distinctive and understudied CSCW shaped problem.
To investigate this space we conducted semi-structured interviews with
archivists and technologists who were directly involved in the selection
of content from the web for archives. These semi-structured interviews
identified thematic areas that inform the appraisal process in web
archives, some of which are encoded in heuristics and algorithms. Making
the infrastructure of web archives legible to the archivist, the
automated agents and the future researcher is presented as a challenge
to the CSCW and archival community.
\end{abstract}

  \section*{ACM Classification Keywords}
  \begin{itemize*}[itemjoin={{; }}, afterlabel={}]
      \item[] H.3.7. Digital Libraries: Systems issues
      \item[] K.4.3. Organizational Impacts: Computer Supported Collaborative Work
    \end{itemize*}


\section{Introduction}\label{introduction}

In 2008, Google estimated that it had 1 trillion unique URLs in its
index {[}2{]}. Recently in 2015, the Internet Archive's homepage
announced that it has archived 438 billion web pages. A simple back of
the envelope calculation indicates that roughly 44\% of the web has been
archived. But this estimate is overly generous. Of course the web has
continued to grow in the 8 years since Google's announcement. The
Internet Archive's count includes multiple snapshots of the same URL
over time. Even Google does not know the full extent of the web, since
much of it is either hidden behind search forms that need to be queried
by humans, the so called deep web {[}54{]}, or blocked from indexing by
Google, the dark web. Consequently, the actual amount of the web that is
archived is not readily available, but certain to be much, much less
than 44\%. Archives of web content matter, because hypertext links are
known to break. Ceglowski {[}11{]} has estimated that about a quarter of
all links break every 7 years. Even within highly curated regions of the
web such as scholarly and legal publishing rates of link rot can be up
to 50\% {[}69,79{]}.

Failing to capture everything should not be surprising to the
experienced archivist. Over the years, archival scholars have argued
that gaps and silences in the archival record are inevitable. This is
partly because we do not have the storage capacity nor all the manpower
nor all the resources required to keep everything. Thus, archivists
necessarily select representative samples, identify unique and
irreplaceable, and culturally valuable, records. We often assume that
archivists abide by a clear set of appraisal principles in their
selection decisions. In practice, selection is a highly subjective
process that reflect the values and aspirations of a privileged few.
More often, archival acquisition also happens more opportunistically and
without adherence to a planned or comprehensive collecting framework.
The central challenge facing the archival community is to better
understand our predisposition to privilege dominant cultures, which
results in gaps in society's archives. As Lyons {[}53{]} recently
argued:

\begin{quote}
If we have any digital dark age, it will manifest, as has been the case
in the past with other forms of information, as a silence within the
archive, as a series of gaping holes where groups of individuals and
communities are absent because there was no path into the archive for
them, where important social events go undocumented because we were not
prepared to act quickly enough, and where new modalities for
communication are not planned for. The digital dark age will only happen
if we, as communities of archives and archivists, do not reimagine
appraisal and selection in light of the historical gaps revealed in
collections today.
\end{quote}

Unlike more traditional archival records the web is a constantly
changing information space. For example, the New York Times homepage
which is uniquely identified by the URL http://www.nytimes.com can
change many times during any given day. In addition, increased
personalization of web content means that there is often not one
canonical version of a particular web document: what one user sees on a
given website can vary greatly compared to what another individual sees.
For instance, what one sees when visiting a particular profile on
Facebook can be quite different from what another person will see,
depending on whether they are both logged in, and part of a particular
network of friends. Even when collecting web content for a particular
institution, such as an academic community, it can be difficult to
discover and delimit relevant regions of content to collect.

Given its vastness, volume of content, and the nature of online media,
capturing and archiving web relies on digital tools. These archiving
tools typically require archivists to supply lists of website URLs or
\emph{seed lists} that are deemed important to capture. These lists are
essentially a series of starting points for a web crawler to start
collecting material. The lists are managed by web archiving software
platforms which then deploy web crawlers or bots that start at a seed
URL and begin to wander outwards into the web by following hyperlinks.
Along with the seeds archivists also supply \emph{scopes} to these
systems that define how far to crawl outwards from that seed URL--since
the limits of a given website can extend outwards into the larger space
of the web, and it is understandably desirable for the bot not to try to
archive the entire web. Different web archiving systems embody different
sets of algorithms and as platforms they offer varying degrees of
insight into their internal operation. In some ways this increasing
reliance on algorithmic systems represents a relinquishing of archival
privilege to automated agents and processes that are responsible for
mundane activity of fetching and storing content. The collaborative
moment in which the archivist and the archival software platform and
agents work together is understudied and of great significance.

The web is an immensely large information space, even within the bounds
a given organization, so given a topic or content area it is often
difficult to even know what website URLs are available, let alone
whether they are relevant or not. The seed list functions as an
interface between the web, the archivist, archival systems, and the
researcher. The seed list also offers material evidence of the
interactions between human and automated agents, and makes the
sociotechnical construction of the web archive legible. It is in this
context that we ask the question: How do technologies for web crawling
and harvesting--seed lists and scopes--figure in their appraisal
decisions?

In this study we focus on answering the research question of how
archivists interact with web archiving systems, and collaborate with
automated agents when deciding what to collect from the web. A better
understanding of this collaborative moment can serve as a foundation for
theories about how archivists are deciding what to collect, that can
then inform the design of web archiving technologies.

\section{Literature Review}\label{literature-review}

The web is very much a part of contemporary life. Archived pieces of the
web will provide essential evidence in the construction of historical
knowledge of our time. Various efforts currently exist to capture the
web at the local, national, international as well as institutional or
community levels {[}50,62,78{]}. Studies of these efforts center around
more efficient methods for capturing the dynamic web and increasing
discoverability of the archived web by examining tools, policies, and
metadata {[}19,38,67{]}.

We have seen reports that point to the limitations of large web
archiving projects. The Wayback Machine for instance cannot capture all
information online and still misses local content, government documents,
and database-backed websites {[}74{]}. Despite its promise to provide
``universal access to all knowledge'' {[}46{]}, cultural heritage
institutions cannot in good faith solely rely on the Internet Archive to
do all the work. Hence, some have advocated for libraries and archives
to take on the responsibility of web archiving in order to capture and
preserve more content. However, the dynamics and mechanics of the
decision-making process over what ends up being archived is not very
much studied.

Deciding what to keep and what gets to be labeled \emph{archival} have
long been a topic of discussion in archival science. Over the past two
centuries archivists have developed a body of research literature around
the concept of \emph{appraisal}, which is the practice of identifying
and retaining records of enduring value. During that time there has been
substantial debate between two leading archival appraisal thinkers,
Hilary Jenkinson {[}44{]} and Theodore Schellenberg {[}70{]}, about the
role of the archivist in appraisal: whether in fact it is the role of
the archivist to decide what records are and are not selected for
preservation {[}34,77{]}. The rapid increase in the amount of records
being generated that began in the mid-20th century, led to the
inevitable (and now obvious) realization that it is impractical and
perhaps impossible to preserve the complete documentary record.
Appraisal decisions must necessarily shape the archive over time, and by
extension our knowledge of the past {[}5,15{]}. It is in the particular
contingencies of the historical moment that the archive is created,
sustained and used {[}8,35,68{]}.

\subsection{Web Archives}\label{web-archives}

The emergence, deployment, and adoption of the World Wide Web has
rapidly accelerated the growth of the documentary record even more,
since its introduction in 1989. The archival community is still in the
process of adapting to this material transformation, and understanding
how it impacts appraisal processes and theories. To put it simply, who
archives what of the web, and how do they do it?

The work of archiving the Internet has been underway since 1996 at the
Internet Archive. Since 2003, the Internet Archive's automated bots have
walked from link to link on the web, archiving what they can along the
way {[}56{]}. Also starting in 2003, organizations belonging to the
International Internet Preservation Consortium (IIPC) began building
their own collections of web content. These collections can either be
country domains like the .uk top-level-domain collected by the British
Library {[}29,63{]}, or specific websites that have been selected
according to a collection development policy. There has been some
attempt at measuring the number and size of web archives {[}30{]}. In
addition, work has gone into investigating the contributions of
automated web archiving agents in the Internet Archive {[}51,52{]}.
There has also been some analysis of how collection development policies
at major national archives enact appraisal {[}60{]}, as well as general
overviews of how archival processes in web archives {[}61{]}. Indeed,
there is no shortage of research material about the need for web
archiving, and technical approaches for achieving it. However, the
practice of appraisal, or how web content is selected for an archive,
which is being performed by a growing number of actors, is currently
understudied.

Surveys of the web archiving community have been conducted by the
International Internet Preservation Consortium (IIPC) {[}55{]} and more
recently by the National Digital Stewardship Alliance (NDSA) in 2011
{[}58{]} and 2013 {[}4{]}. The decade old perspective of the 2004 IIPC
report is now largely anachronistic, but still provides a useful
historical snapshot of web archiving activity at the time, particularly
with regards to the tools used for web archiving. While the fundamental
architecture of the web has been relatively constant, its core standards
(HTTP, HTML, CSS, JavaScript) have been in constant change since then.
The landscape of the web has been profoundly altered by social media
that has allowed millions of individuals around the world to participate
as content creators and to entangle their lived experience with the
infrastructure of the web {[}18{]}. Consequently web archiving tools and
the skills of archivists have required near continuous adaptation in
order to collect and provide access to archived content.

The 2011 and 2013 National Digital Stewardship Alliance (NDSA) surveys
are marked by an attention to the deployment of web archiving
technology. These surveys contribute significantly to an understanding
of how organizations manage, fund and train web archiving teams. It is
notable that the 2013 survey contains the first questions about the
actual content being collected, in terms of its type: social media,
databases, video, audio, blogs, art. How these categories were
determined is not clear. Both the 2011 and 2013 surveys have questions
about inter-institutional, collaborative web archiving, which indicate a
growing interest and engagement with cooperative approaches. However,
specific details about how this cooperation is enacted and achieved are
not detailed.

This lack of detail should not be seen as a failing but rather as a
direct result of the survey method itself, which aims to
representatively sample a population in order to draw statistical
inferences and insights about the larger population of web archiving
efforts. These surveys have been extremely useful for gaining insight
into the activity of a growing international community of web
archivists. Their necessarily broad focus on trends has rationalized and
demarcated the emerging field of web archiving. However, this broad
focus has only provided a very high level view of how particular web
content is selected for archiving. In order to better design systems to
aid in web archiving a thicker description of the day to day work of the
web archivist who appraises content is needed {[}13,23{]}.

The beginnings of such an examination can be found in a series of blog
post interviews with recommending officers at the Library of Congress,
who selected websites for archiving {[}3,32,33,59{]}. These short
vignettes offer a glimpse into how the work of selecting and appraising
web content is achieved in the context of a particular organization.
Using a series of semi-structured interviews Dougherty and Meyer
{[}19{]} closely examined the state of practices in web archives with a
particular focus on the needs of researchers, and the gaps between
research and archival processes. In many ways our study takes a
compatible approach, but instead of identifying ``experts'' in the field
to interview we focused on practitioners who were involved in the day to
day selection of web content for archives.

\subsection{Algorithms and
Infrastructure}\label{algorithms-and-infrastructure}

Thus far, our review has focused on the literatures of archival science
and digital libraries which are essential to positioning our study.
While the study of digital library use and digital libraries as
cyberinfrastructure are no stranger to the CSCW community, there has
been little attention to the specific practice of appraisal in web
archives. Yet we believe there are significant strands of work from CSCW
that have much to offer in understanding how web archives are
constructed--particularly when they are considered not just as
collections of documents, but as sociotechnical systems in which
archivists collaborate with automated agents.

Key among these strands of CSCW work is the study of algorithms and
their role in the construction of bots, platforms, and infrastructures.
Algorithms are increasingly being studied in CSCW contexts that consider
not only the specific workings of the code itself, but also how the
algorithms are components of larger sociotechnical assemblages. Much of
the mundane crawling and downloading of web content in archival systems
is necessarily performed by automated agents, or crawlers, that are
directed in varying degrees by people's interactions with web archiving
software. Much relevant work has been done by Geiger on the use of bots
in Wikipedia to monitor spam and defacement, with specific attention to
the conditions in which code is run {[}24{]}. Geiger has also looked at
collaborative systems for thwarting online harassment in social media
{[}25{]}, which is essentially concerned with content curation on the
web, and closely aligned with archival theory.

The study of algorithms and their social effects is a rapidly growing
area of research which offers multiple modes of analysis for the study
of web archives {[}26{]}. Kitchin provides a review of this literature
while presenting a framework of methodological approaches for the study
of algorithms {[}48{]}. He stresses why thinking critically about
algorithms is so important, which is especially relevant for thinking
about the socio-political dimension of web archives:

\begin{quote}
Just as algorithms are not neutral, impartial expressions of knowledge,
their work is not impassive and apolitical. Algorithms search, collate,
sort, categorise, group, match, analyse, profile, model, simulate,
visualize and regulate people, processes and places. They shape how we
understand the world and they do work in and make the world through
their execution as software, with profound consequences.
\end{quote}

Useful work has also been done examining how users come to understand
the algorithms they interact with. For example how users come to
understand what is visible, and what is invisible, in their social media
feeds {[}22{]}. Seaver {[}71{]} highlights how important cultural
understandings of algorithms are. These various examinations of the
algorithm are highly relevant to analyzing the ways in which archivists
use computational tools while doing appraisal work.

Zooming out a level from algorithms, web archiving tools and systems
themselves can also be understood as software platforms or
socio-political assemblages {[}27,28{]}. Emerging CSCW work that
examines the interactions of policy and governance in the formulation of
software platforms is directly relevant to understanding how web
archiving technologies are situated with respect to intellectual
property, privacy, security and organizational collection development
policies {[}12,39,42{]}.

Finally zooming out even further, we arrive at the perspective of
infrastructure. The CSCW community has had a long and sustained
engagement with research into cyberinfrastructure and infrastructure
studies {[}47,57,64,65{]}. Since its inception in the mid 1990s the
web's emergence as a critical infrastructure has made it an ever present
topic for CSCW work {[}6{]}, especially with regards to collaborative
systems in the sciences. However little attention has been paid to the
specific details of how content \emph{from} the web is archived: who
performs these preservation actions, and what tools and collaborative
systems they use to do it. The CSCW perspective has much to offer since
recognizing and working with breakdowns in web architecture is the
essence archival work, which fits neatly into a lineage of
infrastructure studies beginning with Star {[}72{]} through to the
present day work focused on repair and its impact on design
{[}37,41,66{]}. There has been some engagement with sociotechnical
theory from the field of archival science {[}1,9{]}, however, like CSCW,
it has largely been concerned with the study of scientific data usage,
and not specifically with web infrastructure in general.

This review has covered a great deal of ground in order to make a case
for the study web archives as sociotechnical systems. In order to better
design and evolve appraisal tools for web archives a much richer
understanding of how web content is currently selected by actual
practitioners is needed. We believe that CSCW perspectives, in
particular tried and true ethnographic and practice oriented
methodologies, have a great deal to offer the study web archives
{[}45,49,73{]}. While this theoretical positioning informs our work,
this study is specifically focused on how archivists interact with web
archiving systems as they select material, to gain insight into how
content is selected for preservation. We know that lists of URLs, or
seed lists, are created, since web archiving technologies require them
in order to function. But how URLs end up on these lists is not well
understood. These seed lists are singular artifacts of the intent to
archive, which makes them valuable excavation sites for deepened
understanding of the day to day process of appraisal in web archives.

\section{Methodology}\label{methodology}

To answer our research question we conducted a series of semi-structured
interviews with a carefully selected group of individuals involved in
the selection of web content to explore and excavate these seed lists as
sites of appraisal practice. Rather than providing a statistically
representative or generalizable picture, the goal was to evoke a thick
description of how practitioners enact appraisal in their particular
work environments. Semi-structured interviews were specifically chosen
to add an individualized, humanistic dimension to the existing body of
survey material about the construction of web archives. Archival
appraisal is a socially constructed activity that necessarily involves
the individual archivist, the organization in which they work, and the
broader society and culture in which they live. Consequently the
interviews did not serve as windows onto the appraisal process so much
as they provided insight into how archivists talk about their work with
web archives, specifically with regards to their selection of web
content for archiving or long-term preservation {[}36{]}.

We selected our interview subjects using purposeful sampling that
primarily followed a pattern of stratified sampling, where both typical
cases and extreme cases were selected. Typical cases included
self-identified archivists involved in traditional web archiving
projects at libraries, archives and museums, many of whom were on the
list of attendees at the Web Archives conference in Ann Arbor, Michigan
on November 12-13, 2015. The study also involved participation from
extreme or deviant cases that include individuals who do not necessarily
identify as archivists but are nevertheless involved in web archiving,
such as researchers, local government employees, volunteers, social
activists, and business entrepreneurs. To avoid oversampling from the
users of Archive-It (currently the leading provider of web archiving
services in the United States), we also recruited customers of other web
archiving service providers, such as Hanzo, ArchiveSocial, and members
of the ArchiveTeam community. The organization types identified in the
NDSA survey results {[}4{]} provided a good basis for sampling and
recruitment. However, our own personal familiarity with the small but
growing field of web archiving also informed the development of our
participant list.

In some instances we relied on snowball sampling to recruit interview
participants. There were occasions when the interview subject was not
involved directly in the selection of content for their web archives. In
those cases, we asked if they would refer someone that was more involved
in the actual selection process. Other names were often mentioned during
the interview, and if we felt those individuals could add a useful
dimension to the interview we asked for their contact information.

The study recruited 39 individuals (21 female, 18 male), 27 (13 female,
14 male) of which agreed to be interviewed. A table summarizing the
organization types, occupations, and roles for the interview subjects is
included below. It also includes a designation of whether they were
considered extreme or deviant cases (participants who do not identify
themselves as archivists but are involved in web archiving duties). The
tables illustrate how the our study explicitly focuses on archivists
involved in the selection of web content in a university setting.
Deviant cases such as the role of researcher or developer provide an
opening for future work in this area.

\begin{tabular}{ l c r }
  Organization       &  Deviant  &  n=27 \\
  \hline
  University         &  N        &    19 \\
  Non-Profit         &  Y        &     3 \\
  Museum             &  N        &     2 \\
  Government         &  N        &     1 \\
  Public Library     &  N        &     1 \\
  Activist           &  Y        &     1 \\
\end{tabular}

\begin{tabular}{ l c r }
  Occupation         &  Deviant  &  n=27 \\
  \hline
  Archivist          &  N        &    17 \\
  Manager            &  N        &     7 \\
  Developer          &  Y        &     2 \\
  Professor          &  Y        &     1 \\
\end{tabular}

\begin{tabular}{ l c r }
  Role               &  Deviant  &  n=27 \\
  \hline
  Selector           &  N        &    17 \\
  Technician         &  N        &     6 \\
  Developer          &  Y        &     3 \\
  Researcher         &  Y        &     1 \\
\end{tabular}

Each interview lasted approximately an hour and was allowed to develop
organically as a conversation. The interview protocol guided the
conversation, and provided a set of questions to return to when
necessary. This protocol was particularly useful for getting each
interview started: describing the purpose of the study and the reason
for contacting them. The interview subjects were then asked to describe
their work in web archives, and about their own personal story of how
they had come to that work. After this general introduction and
discussion, the conversation developed by asking follow on questions
about their work and history. The ensuing conversation normally touched
on the interview questions from the protocol in the process of inquiring
about their particular work practices and experiences. Towards the end
of the interview, the interview protocol was also useful in identifying
any areas that had not been covered.

Interviews were conducted via Skype and recorded. Each participant
provided informed consent via email. Participants were located in places
all across the United States, so in person interviews were impractical.
Because of the nature of the study the risks to participants was low,
all interviews were kept confidential and all recordings and transcripts
would be destroyed after the completion of the study. Consequently, we
use pseudonyms to refer to our respondents and the names of their
respective organizations have been obscured.

While this study was not conducted in the field over an extended period
of time, it was deeply informed by ethnographic practices of memoing and
fieldnote taking. These techniques were selected to document the
conversation itself but also to reflect on our involvement and
participation in the web archiving community {[}21{]}. During the remote
interviews memos or jottings were essential for noting particular
moments or insights during the interview. In some cases, these jottings
were useful in highlighting points of interest during the interview
itself. Immediately after each interview, these jottings prompted more
reflective fieldnotes that described notable things that came up in the
interviews. Particular attention was paid to themes that reoccurred from
previous interviews, and new phenomena. As the interviews proceeded, a
file of general reflections helped determine recurring themes and
scenarios as well as unusual cases that encountered.

The process of inductive thematic analysis performed in this study
relied on the use of field notes and personal memos {[}10{]}. The
analysis began by reading all the field notes together, and then
returning to do line by line coding. While coding was done without
reference to an explicit theoretical framework, it was guided by our own
interest in appraisal theory as a sociotechnical system that entangles
the archivist with the material of the web and automated agents.
Interviewee responses that specifically mentioned the selection of
particular web content, and the tools and collaborations they used to
enact that selection were followed up on and explored through open
discussion. This analysis yielded a set of themes that will now be
described.

\section{Findings}\label{findings}

Our study reveals that web archiving involves a variety of technical and
resource constraints that go beyond what is normally considered in
archival appraisal theory. Archival scholars typically characterize
archival selection as a process whereby human actors (archivists)
primarily follow prescribed sets of rules (institutional policies and
professional expectations) to accomplish the task of appraisal and
selection {[}7,15--17,20,31,76{]}. This traditional notion does not
adequately describe how selection occurs in the web archiving context.
Instead, we found that automated agents often serve as collaborators
that act in concert with the archivist. Indeed, these agents themselves
are often the embodiment of rules or heuristics for appraisal. In this
section, we report how crawl modalities, information structures, and
tools play a significant role in selection decisions. We also highlight
how resource constraints as well as moments of breakdown work to shape
appraisal practice.

\subsection{Crawl Modalities}\label{crawl-modalities}

While often guided by archivists in some fashion, the work of archiving
is mostly achieved through a partnership with automated agents (bots)
that do the monotonous, mostly hidden work of collecting web pages. This
work includes fetching the individual HTML for web pages, and then
fetching any referenced CSS, JavaScript, image or video files that are
required for the page to render. Once a given page has been archived the
bot then analyzes the resource for links out into the web, and decides
whether to follow those links to archive them as well. This process
continues recursively until the bot is unable to identify new content
that the archivist has selected, is told to stop, or terminates because
of an unforeseen error. Participants reflected on this process by
talking about the paths that they took, often \emph{with} their
automated agents, through the web in different ways, or modalities:
domain crawls, website crawls, topical crawls, event based crawls,
document crawls.

In domain crawls, a particular DNS name was identified, such as
\emph{example.edu} and the crawler was instructed to fetch all the web
content at that domain. These instructions often included scoping rules
that either allowed the crawler to pull in embedded content from another
domain, such as video content from YouTube, and also to exclude portions
of the domain, such as very large data repositories or so called browser
traps such as calendars that created an infinite space for the crawler
to get lost in.

Website crawls are similar in principle to domain crawls, but rather
than collecting all the content at a particular domain they are focused
on content from a specific DNS host such as \emph{www.example.edu}, or
even a portion of the content made available by that webserver such as
\emph{http://www.example.edu/website/}.

In topic based crawls the archivist was interested in collecting web
material in a specific topical area, such as ``fracking'' or a
particular ``classical music composer''. In order to do this type of
crawl the archivist must first identify the domains or websites for that
topic area. Once a website or domain has been identified the archivist
is able to then instruct the crawler to collect that material.

Event based crawls are similar to topic based crawls but rather than
being oriented around a particular topic they are concerned with an
event anchored at a particular time and place such the Deepwater Horizon
Oil Spill in the Gulf of Mexico. Just as in topic based crawls, host
names or website URLs must be identified first before an automated agent
can be given instructions about what content to collect. With event
based crawls crawling tends to extend over a particular period of time
in which the event is unfolding.

In document crawls the archivist has a known web resource that they want
to collect and add to their archive. This may require an automated agent
of some kind, such as Webrecorder, but also could be a more manual
process where an archivist collects a PDF of a report from a website and
individually deposits it into their archive.

These crawl modalities were often used together, in multiple directions,
by human and non-human agents either working together or separately. For
example in a topical based crawl for fracking related material one
archivist engaged in a discovery process of searching the web using
Google and then following links laterally outwards onto social media
sites and blogs. Once a set of URLs was acquired they were assembled
into a seed list and given to the Archive-It service to collect.

Similarly when an archivist instructed Archive-It to perform a domain
crawl for a large art museum, the resulting dataset was deemed too large
and incomplete. A proliferation of subdomains, and a multiplicity of
content management systems made it difficult to determine the
completeness of the crawl. In this case the archivist used Achive-It's
crawl reports as well as searching/browsing the website to build a list
of particular sub-websites within the domain that were desirable to
archive. Many of these sub-websites were in fact different computer
systems underneath, with their own technical challenges for the web
archiving bot. The larger problem of archiving the entire domain was
made more feasible by focusing on websites discovered doing a failed
domain crawl. This list of websites was then given to Archive-It in the
form of a seed list.

\subsection{Information Structures}\label{information-structures}

In addition to the types of crawling being performed, the activities of
archivists and automated agents were informed by information structures
on and off the web. Primary among these structures encountered in the
interviews were hierarchies, networks, streams and lists. Hierarchies of
information were mentioned many times, but not by name, especially when
an archivist was engaged with collecting the web content of a particular
organization: e.g.~the web content from a particular university or
government agency. This process often involved the use of an
organizational chart or directory listing the components and
subcomponents of the entity in question. One participant talked about
how they used their university's A-Z listing of departments as a way to
build a list of seeds to give to Archive-It. In another example a
government documents librarian used the organizational chart of San
Mateo local government to locate web properties that were in need of
archiving.

Not all web archiving projects are fortunate enough to have an explicit
hierarchical map. Many appraisal activities involve interacting with and
discovering networks of resources, that extend and cut across across
organizational and individual boundaries. For example when Vanessa was
archiving web content related to the Occupy social movement she saw her
organization's interest in collecting this content fold into her own
participation as an activist. This enfolding of interest and
participation was evident in the network structure of the web where her
personal social media connections connected her to potential donors of
web content.

\begin{quote}
It was part of that same sort of ecosystem of networks. It became clear
to me through that process how important that network is becoming in
collecting social movements moving forward. It was interesting watching
people who had been doing collecting for decades in activist networks
that they were a part of, and then these new activist
networks\ldots{}there wasn't a whole lot of overlap between them, and
where there was overlap there was often tension. Unions really wanted in
on Occupy and young people were a little bit wary of that. So social
media networks became really important.
\end{quote}

In another example a network of vendor supplied art agents supplied a
museum with gallery catalogs, which were then used to identify gallery
and artist websites of research value. Physical networks of agents,
artists and galleries undergirded the networks of discovered websites.
Indeed this particular museum used multiple vendors to perform this
activity.

Another information structure that participants described as part of
their appraisal process was ``information streams.'' Information streams
are content flows on the web that can be tapped into and used for the
selection of content for a web archive. For example Roger who worked for
a non-profit web archive described how they developed a piece of
software to use a sample of the Twitter firehose to identify web
resources that are being discussed. Roger also described how edit
activity on Wikipedia involving the addition of external links was used
to identify web resources in need of archiving. Nelson who worked as a
software developer for another volunteer organization described how he
used RSS feeds to identify new news content that was in need of
archiving. More traditional streams of content in the form of mailing
lists and local radio, pushed content to several archivists. These
streams were analyzed for reference to people, organizations and
documents to seek out on the web.

While they are a bit more abstract, participants also described
interacting with lists of information. The most common example of this
was lists of URLs in the host reports from the Archive-It service which
allowed archivists to review what host names were and were not present
in their capture of web content. For example Dorothy who was collecting
her university's domain:

\begin{quote}
I definitely remember there was a lot of trial and error. Because
there's kind of two parts. One of them is blocking all those extraneous
URLs, and there were also a lot of URLs that are on the example.edu
domain that are basically junk. Like when sometimes Archive-It hits a
calendar it goes into an infinite loop trying to grab all the pages from
the calendar. So what I would typically do is look at the list of the
URLs. Once you've done a crawl, a real crawl, or a test crawl that
doesn't actually capture any data, there's this report that has a list
of hosts, for example facebook.com, twitter.com and then next to that
there's a column called URLs and if you click the link you get a file,
or if it's small enough a web page that lists all the URLs on that
domain. So one thing that I would try to do is visually inspect the list
and notice if there's a lot of junk URLs.
\end{quote}

The question of what is and what is not junk is the central question
facing the archivist when they attempt to archive the web. The reports
that Archive-It provides at the host name level are an indicator of
whether the crawl is missing or including things that it should not.
Scanning lists of host names and URLs happened iteratively as multiple
crawls were performed.

When considering how participants talked about these hierarchies,
networks, streams and lists of information it became clear that they
were traversing these structures themselves using their browser, as well
as instructing and helping the archival bots do the same. The domain
knowledge of the archivist was a necessary component in this activity,
as was the ability for the bot to rapidly perform and report on highly
repetitive tasks.

\subsection{Time and Money}\label{time-and-money}

Another thematic feature that emerged from the fieldnotes around the
interviews were the material constraints of time and money in the
human-machine collaboration of web archiving. Time and money are
combined here because of the way they abstract, commensurate and make
appraisal practices legible.

Many web archiving projects cited the importance of grant money in
establishing web archiving programs. These grants often were focused on
building technical capacity for web archiving, which itself is not
directly tied to the appraisal process. However it is clear that the
technical ability to archive web content is a key ingredient to
performing it. Grant money was also used to archive particular types of
web content. For example, one university used grant money to archive
music related web content, and another university received a grant to
focus on state government resources.

The most common way that money was talked about by participants was in
subscription fees for web archiving services. Archive-It is a web
archiving service where organizations pay an annual subscription fee to
archive web content. The primary metric of payment is the amount of data
collected in a given year. Interviewees often mentioned that their
ability to crawl content was informed by their storage budget. In one
example an archivist set the scoping rules for a full domain crawl of
her university such that software version control systems were ignored
because of the impact it was having on their storage allocation.
Dorothy, who was a user of the ArchiveSocial service needed to reduce
the number of local government social media accounts that it was
archiving because her subscription only allowed a certain number of
accounts to be collected.

Time manifested in the appraisal of web content at human and machine
scales. In one common pattern, archivists set aside time every week, be
it a day, or a few hours, for work on the discovery of web content. In
one case, Wendy set aside time to read filtered emails about local news
stories. In Lisa's case, she set aside a meeting time every week for her
acquisition team to get together and review potential web sites for
archiving by inspecting websites together on a large screen monitor.

Time was also evident in the functioning of automated agents, because
their activity was often constrained and parameterized by time. For
example archivists talked about running test crawls in Archive-It for no
longer than 3 days. Dorothy talked about the information being gathered
in near real time from social media accounts that Archive-It was
monitoring:

\begin{quote}
The archiving is by the minute. So if I post something, and then edit it
in five minutes then it is archived again. If someone comments on
something and then another person comments it is archived again. You
don't miss anything. A lot of the other archiving companies that we've
talked to say they archive a certain number of times a day: maybe they
archive at noon, and at 5, and at midnight, and there's an opportunity
to miss things that people deleted or hid.
\end{quote}

In this case the software was always on, or at least appeared to be
always on at human time scales. The web content itself also had a time
dimension that affected appraisal decisions. For example the perceived
cumulativeness of a website was an indicator of whether or how often
material was in need of archiving. Blogs, in particular, were given as
examples of websites that might need to be crawled less because of the
ways that they accumulated, and did not remove content.

Another motivation for linking time and money in this way is because of
how they entail each other. The time spent by archivists in discovery
and evaluation of web content for archiving often has a monetary value
in terms of salary or hourly wages. Similarly the amount of time spent
crawling is often a function of the amount of data acquired, and the
cost for storage.

\subsection{People}\label{people}

One might assume that the work to appraise web archives necessarily
involves archivists. However, our interview data made it clear that not
all the people involved in appraisal called themselves archivists, and
they often worked together with human and non-human agents in
collaborative relationships that extended beyond the archives itself.

At one large university archives, a series of individuals were involved
in the establishment of their web archives. Their effort extended over a
15-year period that started with Kate who pioneered the initial work
that ultimately led to a mainstreaming of web content into the archives.
Multiple staff members, including Jack and Deb who were field archivists
responsible for outreach into the university community, and around the
state. The field archivists selected web content, which was communicated
to Phillip, another archivist, who managed their Archive-It
subscription, performed crawls and quality assurance. Jack and Deb
actively sought out records in their communities by interviewing
potential donors, to determine what types of physical and electronic
records were valuable.

John worked as a software developer for a volunteer organization that
performed focused collecting of web content that was in danger of being
removed from the web. He was a physics student who was interested in
using his software development skills to help save at risk web content.
John collaborated with 20-30 other volunteers, one of whom is Jane who
worked at a large public web archive, and was routinely contacted via
email and social media when websites were in danger of disappearing.

Many interviewees reflected on their own participation in the activities
and events that they were documenting. Recall Vanessa who was working to
collect web content related to the Occupy movement. She and her
colleagues at the library worked to document the meetings and protests
from within the movement itself. One of her colleagues worked on the
minutes working group which recorded and made available the proceedings
of the meetings. In another case two archivists and separate
institutions were working together to document the use of fracking in
their respective geographic areas. They worked together to partition the
space as best they could by region, but many businesses and activist
organizations worked across the regions.

While collaboration across organizational boundaries was evident,
several participants noted that duplication of web content was not
widely viewed as something to be avoided. Many commented that
duplication was one way to ensure preservation, following ``lots of
copies keeps stuff safe'' (LOCKSS). Local copies of resources that are
available elsewhere can be of benefit when using the data:

\begin{quote}
If I can't get a copy it doesn't exist in the same way. I think that
there is still a lot to being able to locally curate and manage
collections and the fact that it's over in another space limits, or puts
some limits on the things that can be done with the data now and in the
future. Sure right now I've got a great relationship with a guy that
knows how to get the stuff. But what happens in five years when those
relationships end? How do our students and researchers get access to the
data then?
\end{quote}

In addition the locus of web archiving work shifted within organizations
from one department to the other as key individuals left the library,
and as web content was migrated from one system to another. This
turbulence was common, especially in the use of fellowships and other
temporary positions.

\subsection{Tools}\label{tools}

We have already discussed some tools of the trade that archivists use
for collecting websites: the Internet Archive, Archive-It, ArchiveSocial
and Hanzo are notable ones that came up during the interviews. These
tools are really more like services, or assemblages of individual tools
and people interacting in complex and multilayered ways. An
investigation of each of these services could be a research study in
themselves. These tools largely require intervention by a person who
guides the tool to archive a particular website, or set of web resources
using a seed list or the equivalent. Rather than dig into the particular
systems themselves it is useful to attend to the ways which tools were
used to fill in the gaps between these platforms and their users.

Consider the ways in which spreadsheets were used almost ubiquitously by
interviewees. These spreadsheets were occasionally used by individuals
in relative isolation, but were most often used to collaboratively
collect potential websites that were of interest. Google Sheets in
particular allowed individuals to share lists of URLs and information
about the websites. Archivists would share read-only or edit level
permissions for their spreadsheets to let each other know what was being
collected. These spreadsheets were later transferred into a web
archiving service like Archive-It as seed lists. In the process much of
the additional information, or provenance metadata concerned with the
selection of a website was lost in translation.

Often times web forms of various kinds were used as front ends on these
spreadsheets. These forms mediated access to the spreadsheets and
provided additional guidance on what sorts of data were required for
nominating a web resource for the archive. Tracy developed a custom
application for tracking nominations, so different parties could see
each other's nominations. Tracy noted that one of its drawbacks was that
the tool did not link to the archived web content when it was acquired.

Email was also widely used as a communication tool between selectors of
websites and the individuals performing the web crawling. In one case a
technician would receive requests to crawl a particular website via
email, which would initiate a conversation between the technician and
the selector to determine what parts of the website to archive. This
process would often involve the technician in running test crawls to see
what problem areas there were. Several archivists spoke about how they
subscribed to specific local news aggregators that collected news
stories of interest.

However, email was not the only communication method used in the
appraisal process. As already noted social media, particularly Twitter,
was used as a way of communicating with prominent web archiving
individuals when websites were in need of archiving. In one case IRC
chat was also a way for volunteers to talk about websites that were in
need of archiving, and to coordinate work. These conversations were
extremely important because they embody the process of determining the
value of resources.

Many interviewees used the Archive-It service and commented on the
utility of test crawls. Test crawls were essentially experiments where
the archivist instructed the crawler to archive a particular URL using
particular scoping URLs to control how far the crawl proceeded. Once the
crawl was completed the archivist would examine the results by browsing
the content and comparing to the live website. The archivist would also
examine reports to look at the amount of data used, URLs that were
discovered but not crawled either because of time or because they were
blocked by the scope rules. The experiments were iterative in that the
results of one test would often lead to another refined test until the
crawl was deemed good. Almost all participants talked about this process
as quality assurance or QA instead of appraisal, despite the fact that
it was ultimately a question of what would and would not go into the
archive. One exception to that rule was an archivist who had 10 years
experience doing web archiving with multiple systems who referred to
this as pre-crawl and post-crawl appraisal.

It is notable to observe how engineering terminology like quality
assurance has crept into the language of the archive where appraisal
would be a more apt term. One archivist also noted how archival notions
of processing and appraisal which are normally thought of as distinct
archival activities get folded together or entangled in the process of
test crawling. Indeed one participant went so far as to say that the
process of web archiving actually felt more like collection building
than archiving.

In few cases, the Domain Name Service itself was used as a service to
discover subdomains that were part of a university's domain. A large
number of target hostnames were discovered, which were then prioritized
in order to build a seed list. In another case knowledge of the rules
around the .mil DNS top level domain were used to determine websites of
interest for archiving government sites. However these rules were
imperfect as some US government websites would use the .com top level
domain, such as US Postal Service.

Another prominent technology that participants mentioned was content
management systems. In many cases archivists had experience working as
web designers or information architects. They had used content
management systems like Drupal, Ruby on Rails, WordPress, etc. The
archivist would use this knowledge to decide how to crawl websites and
diagnose problems when they arose.

\subsection{Breakdown}\label{breakdown}

One of the more salient findings during analysis was the locus of
breakdown which made the relations between people, tools, and web
infrastructure more legible. These moments of breakdown also lead to
greater understanding of how the tools operated, and generated
opportunities for repair and innovation {[}40{]}.

Charles was attempting to do a full domain crawl of his university's
domain with the Archive-It tool. An unfortunate side effect of running
this crawl was that portions of the university website were put under
more significant load than usual, became unresponsive, and crashed. IT
specialists that were responsible for these systems incorrectly
identified the crawlers as a denial of service attack, and traced them
to Archive-It. An email conversation between the technicians and
Archive-It led to the technicians at the university connecting up with
the archivists who were attempting to archive web content--at the same
institution. This situation led to lines of communication being opened
between the library and the central IT which were not previously
available. It also led to increased understanding of the server
infrastructure at the university which was housed in four different
locations. The IT department became aware of the efforts to archive the
university's web spaces, and began to notify the archivist when
particular websites were going to be redesigned or shutdown and in need
of archiving.

In another case John used a command line web crawling and archiving tool
called wget to collect web content. wget was used to generated a
snapshot of web content and serialize it using the WARC file format. He
then used another piece of software playback tool called
WebarchivePlayer to examine the data stored in the WARC file to see how
complete the archive was. In some cases he would notice missing files or
content that failed to load because the browser was attempting to go out
to the live web and he had disabled Internet access. This breakdown in
the visual presentation of web resources would prompt John to use the
browser's developer tools to look for failed HTTP requests, and trace
these back to JavaScript code that was dynamically attempting to collect
content from the live web. He would then also use this knowledge to
craft additional rules for wget using the Lua programming language, to
fetch the missing resources. When his examination of the WARC file
yielded a satisfactory result the resulting Lua code and wget
instructions were bundled up and deployed to a network of crawlers that
collaborated to collect the website.

As previously discussed, storage costs are another point of breakdown
when archivists are deciding what web content to archive. Several
participants mentioned their use of test crawls in an attempt to gauge
the size of a website. The full contours of a website are difficult to
estimate, which makes estimating storage costs difficult as well. Some
participants were able to communicate with individuals who ran the
website being archived in order to determine what content to collect.
Roger, who was mentioned earlier, was able to got into conversation with
an engineer who worked at a video streaming service which was in the
process of being sold. Together they determined that the full set of
data was 1.1 petabytes in size, which (after consultation with the
directory of that archive) made it very difficult to think about
archiving in full.

\begin{quote}
I went back to the developer and asked: could you give me a tally of how
many videos have had 10 views, how many videos have had 100 views and
how many videos have had a 1000 views? It turned out that the amount of
videos that had 10 views or more was like 50-75 TB. And he told me that
50\% of the videos, that is to say 500 TB had never been viewed. They
had been absorbed and then never watched. A small amount had been
watched when they were broadcast and never seen again. We had to walk
away from the vast majority. Given that we can't take them all, what are
the most culturally relevant ones? We grabbed mostly everything that was
10 or more. The debate is understandable. In an ideal world you'd take
it all. The criteria we've tended to use is, I always like to grab the
most popular things, and the first things. So if you have a video
uploading site I want the first year of videos. I want to know what
people did for the first year when they were faced with this because
there's no questions this is very historically relevant. But I also want
people to have what were the big names, what were the big things that
happened. And that's not perfect.
\end{quote}

In this case a breakdown that resulted from the size of the collection
and the available storage became a site for innovation, and an
opportunity to make legible appraisal decisions around what constitutes
culturally significant material.

Another extremely common case of breakdown is when a robots.txt file
prevented a crawler from creating a high fidelity capture of a website.
A robots.txt file instructs automated agents in what resources it can
and cannot request. Frequently content management systems will block
access to CSS or image files which makes a web archive of the pages
visibly incomplete, and difficult to use. Many (but not all) archives
attempted to be polite by instructing their web archiving bots to
respect these robots.txt files. When they encountered a problem they
would often need to reach out to someone at the organization hosting the
website. When contact was made the robots.txt file would sometimes be
adjusted to allow the bot in. The archivist became aware of how the
website was operating and the website owner became aware of the
archiving service. In one instance this communication channel led a
website owner to make more cumulative information available on their
website instead of replacing (and thus removing) older content. In some
sense the website itself adapted or evolved an archival function based
on the interactions between the archivist and the manager of the website
being archived.

\section{Discussion and Future Work}\label{discussion-and-future-work}

On the one hand these research findings demonstrate a somewhat mundane
but perhaps comforting finding that in many ways appraisal processes in
web archives appear to be congruent with traditional notions of
appraisal. We saw documentation strategies {[}68{]} at play in many
cases where a collaboration between records creators, archives and their
users informed decisions about what needed to be collected from the web.
The functional analysis appraisal technique was also used by archivists
as they analyzed the structure of organizations in order to determine
what needed to be collected. We also saw postcustodial theory {[}14{]}
in operation when archivists interacted with website owners, and in some
cases encouraged them to adopt archival practices. So rather than a
particular archival institution being responsible for the preservation
and access to documents, the responsibility is spread outwards into the
community of web publishing.

A recurring theme in the analysis above was the archivists' attention to
contemporary culture and news sources. We recall one participant who
spoke of her mentor, who had set an example of taking two days every
week to pore over a stack of local newspapers, and clip stories that
contained references to local events, people and organizations to
explore as record sources. She spoke of how she continued this tradition
by listening to local radio, subscribing to podcasts, RSS feeds and
email discussion lists. She then regularly noted names of organizations,
people and events in these streams as potential record sources. While
not all interviewees spoke explicitly of this practice being handed
down, the attention to local news sources was a common theme,
particularly when it came to processing information streams. This
attention to current events while simple, is extraordinarily powerful,
and reminiscent of Booms {[}8{]}:

\begin{quote}
The documentary heritage should be formed according to an archival
conception, historically assessed, which reflects the consciousness of
the particular period for which the archives is responsible and from
which the source material to be appraised is taken. (p.~105)
\end{quote}

Echoes of Booms can be found in this description by Roger of how his
archive's appraisal policies are enacted:

\begin{quote}
The greater vision, as I interpret it, is that we allow the drive of
human culture to determine what is saved. Not to the exclusion of
others, but one really good source of where things are that need to be
saved is to see what human beings are conversing about and what they are
interacting with online.
\end{quote}

Websites, search engines and social media platforms are material
expressions of the transformation of content into computational
resources, with centers of power and influence that are new, but in many
ways all too familiar. The continued challenge for archivists is to tap
into these sources of information, to deconstruct, and reconstruct them
in order to document society, as Booms urged. In the shift to
computational resources there is an opportunity to design systems that
make these collaborations between archivists, automated agents and the
web legible and more understandable for all parties, and particular for
the future researcher who is trying to understand why something has been
archived.

The appraisal processes that are being enacted by archivists are not
always adequately represented in the archive itself. Recall the
spreadsheets, emails and chat systems that are used during appraisal,
that all but disappear from the documentary record. These systems are
being used to fill the broken spaces or gaps in the infrastructure of
web archives. Each of these hacks, or attempts at creatively patching
archival technology, is a design hint that can inform the affordances
offered by archival tools and platforms. For example, if spreadsheets
can be collaboratively used by a group of archivists to record why a web
resource was selected, who selected it, and other administrative notes,
perhaps this collaborative functionality could be incorporated into the
web archiving platforms themselves? One opportunity of future work would
be to examine these sites of breakdown in greater detail, in order to
help archivists and their automated agents create a more usable and
legible archival record. Further examination of how consensus is
established when archivists are collaborating would also be a fruitful
area to explore in order to understand \emph{how} archivists are
collaborating with each other using these technical systems.

Another significant theme that points to an area for future research is
found in the collaborative sociotechnical environment made up of
archivists, researchers (the users of the archive), and the
systems/tools they use in their work. The inner workings of the archive
always reflect or reinscribe the media they attempt to preserve and
provide access to. As electronic records and the World Wide Web have
come to predominate, the architecture of the archive itself has
necessarily been transformed into a computerized, distributed system,
whose data flows and algorithms reshape the archival imagination itself
{[}75{]}. Even with its narrow focus on the appraisal decisions made by
archivists this study demonstrates that archivists have rich and highly
purposeful interactions with algorithmic systems as they do their work
of selecting and processing web resources. Time and again we saw that
archivists used these systems, and cleverly arrived at techniques for
understanding the dimensions of the resources they were collecting, the
fidelity of the representations created, and ultimately the algorithmic
processes that they were directing.

As outlined by Kitchin {[}48{]} the study of these \emph{archival
algorithmic systems} offers several fruitful areas for future work, the
most promising of which could be the: the study of the algorithms
themselves (their source code) in conjunction with an ethnographic
analysis of the environments where those platforms are developed. This
work could study a particular platform such as Archive-It or the
operations of a specific group such as ArchiveTeam or the British
Library in order to better understand how these archival systems are
designed and used. Attention to how policy and governance concerns
become entangled with the design of these preservation systems would be
also be extremely valuable {[}39{]}. In addition it could be useful to
turn the analysis inward and examine the practices of non-archivists,
such as CSCW researchers themselves, as they collect and manage content
retrieved from the web. How do their practices compare to those of the
emerging web archiving professional?

\section{Conclusion}\label{conclusion}

Take a moment to imagine a science fiction future where the archival
record is complete. Nothing is lost. Everything is remembered. To some
this is a big data panacea and to others a dystopian nightmare of the
panopticon. Fortunately we find ourselves somewhere in between these two
unlikely extremes. The archive is always materially situated in society.
This study asked a simple question of how URLs end up being selected for
an archive. The responses illustrate that archivists, bots, record
creators and the web operate in a flattened space, where it is often
difficult to assess where one begins and the other ends {[}43{]}. They
also highlighted that much work remains to be done to understand how web
archives operate as a sociotechnical system, and how that understanding
can inform the design of better tools.

Our ability to collect and preserve records is a function not only of
technology but our laws, values and ethics as well as the resources at
hand to make them real. The web is not vastly different. But what is
different is that the material of the web and our computational
infrastructures provide us with new opportunities to make legible the
values and ethics that are inscribed in our decisions of what to keep,
and what not to keep. Are we up to the challenge?

\section{Acknowledgements}\label{acknowledgements}

We would like to thank Samantha Abrams, Nicholas Taylor, Kari Kraus,
Leah Findlater and Jessica Vitak for their valuable feedback and
guidance during the development and implementation of this study. Also,
a big thanks to members of the Web archiving community who we
interviewed. You all displayed such generosity, enthusiasm and care when
talking about your work.

\section*{References}\label{references}
\addcontentsline{toc}{section}{References}

\hyperdef{}{ref-Akmon:2011}{\label{ref-Akmon:2011}}
1. Dharma Akmon, Ann Zimmerman, Morgan Daniels, and Margaret Hedstrom.
2011. The application of archival concepts to a data-intensive
environment: Working with scientists to understand data management and
preservation needs. \emph{Archival Science} 11, 3: 329--348.
\href{http://doi.org/10.1007/s10502-011-9151-4}{https://doi.org/10.1007/s10502-011-9151-4}

\hyperdef{}{ref-Alpert:2008}{\label{ref-Alpert:2008}}
2. Jesse Alpert and Nissan Hajaj. 2008. We knew the web was big.
Retrieved from
\url{https://googleblog.blogspot.com/2008/07/we-knew-web-was-big.html}

\hyperdef{}{ref-Anderson:2011}{\label{ref-Anderson:2011}}
3. Kimberly D Anderson. 2011. Appraisal learning networks: How
university archivists learn to appraise through social interaction.
University of California, Los Angeles.

\hyperdef{}{ref-Bailey:2013}{\label{ref-Bailey:2013}}
4. Jefferson Bailey, Abigail Grotke, Kristine Hanna, Cathy Hartman,
Edward McCain, Christie Moffatt, and Nicholas Taylor. 2013. \emph{Web
archiving in the United States: A 2013 survey}. National Digital
Stewardship Alliance. Retrieved from
\url{http://digitalpreservation.gov/ndsa/working_groups/documents/NDSA_USWebArchivingSurvey_2013.pdf}

\hyperdef{}{ref-Bearman:1989}{\label{ref-Bearman:1989}}
5. David Bearman. 1989. Archival methods. \emph{Archives and Museum
Informatics} 3, 1. Retrieved from
\url{http://www.archimuse.com/publishing/archival_methods/}

\hyperdef{}{ref-Bentley:1997}{\label{ref-Bentley:1997}}
6. Richard Bentley, Thilo Horstmann, and Jonathan Trevor. 1997. The
World Wide Web as enabling technology for CSCW: The case of BSCW. In
Richard Bentley, Uwe Busbach, David Kerr and Klaas Sikkel (eds.),
\emph{Groupware and the World Wide Web}. Springer, 1--24.

\hyperdef{}{ref-Boles:1985}{\label{ref-Boles:1985}}
7. Frank Boles and Julia Young. 1985. Exploring the black box: The
appraisal of university administrative records. \emph{The American
Archivist} 48, 2: 121--140.

\hyperdef{}{ref-Booms:1987}{\label{ref-Booms:1987}}
8. Hans Booms. 1987. Society and the formation of a documentary
heritage: Issues in the appraisal of archival sources. \emph{Archivaria}
24, 3: 69--107. Retrieved from
\url{http://journals.sfu.ca/archivar/index.php/archivaria/article/view/11415/12357}

\hyperdef{}{ref-Botticelli:2000}{\label{ref-Botticelli:2000}}
9. Peter Botticelli. 2000. Records appraisal in network organizations.
\emph{Archivaria} 1, 49.

\hyperdef{}{ref-Braun:2006}{\label{ref-Braun:2006}}
10. Virginia Braun and Victoria Clarke. 2006. Using thematic analysis in
psychology. \emph{Qualitative research in psychology} 3, 2: 77--101.

\hyperdef{}{ref-Ceglowski:2011}{\label{ref-Ceglowski:2011}}
11. Maciej Ceglowski. 2011. Remembrance of links past. Retrieved from
\url{https://blog.pinboard.in/2011/05/remembrance_of_links_past/}

\hyperdef{}{ref-Centivany:2016}{\label{ref-Centivany:2016}}
12. Alissa Centivany. 2016. Policy as embedded generativity: A case
study of the emergence and evolution of hathiTrust. In \emph{Proceedings
of the 19th ACM Conference on Computer-Supported Cooperative Work \&
Social Computing}, 924--938.

\hyperdef{}{ref-DeCerteau:2011}{\label{ref-DeCerteau:2011}}
13. Michel de Certeau. 2011. \emph{The practice of everyday life}.
University of California Press.

\hyperdef{}{ref-Cook:1993}{\label{ref-Cook:1993}}
14. Terry Cook. 1993. The concept of the archival fonds in the
post-custodial era: Theory, problems and solutions. \emph{Archivaria}
35: 24--37.

\hyperdef{}{ref-Cook:2011}{\label{ref-Cook:2011}}
15. Terry Cook. 2011. We are what we keep; we keep what we are: Archival
appraisal past, present and future. \emph{Journal of the Society of
Archivists} 32, 2: 173--189.

\hyperdef{}{ref-Couture:2005}{\label{ref-Couture:2005}}
16. Carol Couture. 2005. Archival appraisal: A status report.
\emph{Archivaria} 1, 59: 83--107.

\hyperdef{}{ref-Cox:2010}{\label{ref-Cox:2010}}
17. Richard J Cox. 2010. Archivists and Collecting. In Marcia Bates and
Mary Niles Maack (eds.), \emph{Encyclopedia of library and information
sciences} (3rd ed.). Taylor \& Francis, 208--220.

\hyperdef{}{ref-Dijk:2013}{\label{ref-Dijk:2013}}
18. José van Dijk. 2013. \emph{The culture of connectivity: A critical
history of social media}. Oxford University Press. Retrieved from
\url{https://global.oup.com/academic/product/the-culture-of-connectivity-9780199970780}

\hyperdef{}{ref-Dougherty:2014}{\label{ref-Dougherty:2014}}
19. Meghan Dougherty and Eric T Meyer. 2014. Community, tools, and
practices in web archiving: The state-of-the-art in relation to social
science and humanities research needs. \emph{Journal of the Association
for Information Science and Technology} 65, 11: 2195--2209.

\hyperdef{}{ref-Eastwood:1992}{\label{ref-Eastwood:1992}}
20. Terry Eastwood. 1992. Towards a social theory of appraisal.
\emph{The archival imagination: essays in honour of Hugh A. Taylor}:
71--89.

\hyperdef{}{ref-Emerson:2011}{\label{ref-Emerson:2011}}
21. Robert M Emerson, Rachel I Fretz, and Linda L Shaw. 2011.
\emph{Writing ethnographic fieldnotes}. University of Chicago Press.

\hyperdef{}{ref-Eslami:2016}{\label{ref-Eslami:2016}}
22. Motahhare Eslami, Karrie Karahalios, Christian Sandvigt, Kristen
Vaccaro, Aimee Rickman, Kevin Hamilton, and Alex Kirlik. 2016. First i
like it, then i hide it: Folk theories of social feeds. In
\emph{Proceedings of the 2016 cHI conference on human factors in
computing systems}, 2371--2382. Retrieved from
\url{http://www-personal.umich.edu/~csandvig/research/Eslami_FolkTheories_CHI16.pdf}

\hyperdef{}{ref-Geertz:1973a}{\label{ref-Geertz:1973a}}
23. Clifford Geertz. 1973. Thick Description: Toward an Interpretive
Theory of Culture. In \emph{The interpretation of cultures: Selected
essays}. Basic books.

\hyperdef{}{ref-Geiger:2014}{\label{ref-Geiger:2014}}
24. R. Stuart Geiger. 2014. Bots, bespoke, code and the materiality of
software platforms. \emph{Information, Communication \& Society}:
342--356. Retrieved from
\url{http://www.tandfonline.com/doi/full/10.1080/1369118X.2013.873069}

\hyperdef{}{ref-Geiger:2016}{\label{ref-Geiger:2016}}
25. R. Stuart Geiger. 2016. Administrative support bots in Wikipedia:
How algorithmically-supported automation can transform the affordances
of platforms and the governance of communities.

\hyperdef{}{ref-Gillespie:2015}{\label{ref-Gillespie:2015}}
26. Tarleton Gillespie and Nick Seaver. 2015. Critical algorithm
studies: A reading list. Retrieved from
\url{https://socialmediacollective.org/reading-lists/critical-algorithm-studies/}

\hyperdef{}{ref-Gillespie:2010}{\label{ref-Gillespie:2010}}
27. Tarleton Gillespie. 2010. The politics of platforms. \emph{New
Media \& Society} 12, 3: 347--364.

\hyperdef{}{ref-Gillespie:2017}{\label{ref-Gillespie:2017}}
28. Tarleton Gillespie. 2017. Governance of and by platforms. In Jean
Burgess, Thomas Poell and Alice Marwick (eds.), \emph{Sage handbook of
social media}. Sage.

\hyperdef{}{ref-Gomes:2006}{\label{ref-Gomes:2006}}
29. Daniel Gomes, Sérgio Freitas, and Mário J Silva. 2006. In
\emph{Design and selection criteria for a national web archive}.
Springer, 196--207.

\hyperdef{}{ref-Gomes:2011}{\label{ref-Gomes:2011}}
30. Daniel Gomes, João Miranda, and Miguel Costa. 2011. In \emph{A
survey on web archiving initiatives}. Springer, 408--420.

\hyperdef{}{ref-Greene:1998}{\label{ref-Greene:1998}}
31. Mark Greene. 1998. ``The surest proof'': A utilitarian approach to
appraisal. \emph{Archivaria}: 127--169.

\hyperdef{}{ref-Grotke:2011}{\label{ref-Grotke:2011}}
32. Abbie Grotke. 2011. Ask the recommending officer: The civil war
sesquicentennial web archive. Retrievedfrom
\url{http://blogs.loc.gov/digitalpreservation/2011/08/ask-the-recommending-officer-the-civil-war-sesquicentennial-web-archive/}

\hyperdef{}{ref-Grotke:2012}{\label{ref-Grotke:2012}}
33. Abbie Grotke. 2012. Ask the recommending officer: Indian general
elections 2009 web archive. Retrievedfrom
\url{http://blogs.loc.gov/digitalpreservation/2012/01/ask-the-recommending-officer-indian-general-elections-2009-web-archive/}

\hyperdef{}{ref-Ham:1993}{\label{ref-Ham:1993}}
34. F Gerald Ham. 1993. \emph{Selecting and appraising archives and
manuscripts}. Society of Amer Archivists. Retrieved from
\url{http://catalog.hathitrust.org/Record/002650426}

\hyperdef{}{ref-Harris:2002}{\label{ref-Harris:2002}}
35. Verne Harris. 2002. The archival sliver: Power, memory, and archives
in South Africa. \emph{Archival Science} 2, 1-2: 63--86.

\hyperdef{}{ref-Holstein:2011}{\label{ref-Holstein:2011}}
36. James A Holstein and Jaber F Gubrium. 2011. Animating interview
narratives. \emph{Qualitative research} 3: 149--167.

\hyperdef{}{ref-Houston:2016}{\label{ref-Houston:2016}}
37. Lara Houston, Steven J Jackson, Daniela K Rosner, Syed Ishtiaque
Ahmed, Meg Young, and Laewoo Kang. 2016. Values in repair. In
\emph{Proceedings of the 2016 CHI conference on human factors in
computing systems}, 1403--1414.

\hyperdef{}{ref-Huurdeman:2015}{\label{ref-Huurdeman:2015}}
38. Hugo C Huurdeman, Jaap Kamps, Thaer Samar, Arjen P de Vries, Anat
Ben-David, and Richard A Rogers. 2015. Lost but not forgotten: Finding
pages on the unarchived web. \emph{International Journal on Digital
Libraries} 16, 3-4: 247--265.

\hyperdef{}{ref-Jackson:2014c}{\label{ref-Jackson:2014c}}
39. Steven J Jackson, Tarleton Gillespie, and Sandy Payette. 2014. The
policy knot: Re-integrating policy, practice and design in CSCW studies
of social computing. In \emph{Proceedings of the 17th ACM Conference on
Computer Supported Cooperative Work \& Social Computing}, 588--602.

\hyperdef{}{ref-Jackson:2014}{\label{ref-Jackson:2014}}
40. Steven J. Jackson. 2014. Rethinking Repair. In Pablo Boczkowski and
Kirsten Foot (eds.), \emph{Media technologies: Essays on communication,
materiality and society}. MIT Press. Retrieved from
\url{http://sjackson.infosci.cornell.edu/RethinkingRepairPROOFS(reduced)Aug2013.pdf}

\hyperdef{}{ref-Jackson:2014a}{\label{ref-Jackson:2014a}}
41. Steven J. Jackson and Laewoo Kang. 2014. Breakdown, obsolescence and
reuse: HCI and the art of repair. Retrieved from
\url{http://sjackson.infosci.cornell.edu/Jackson\&Kang_BreakdownObsolescenceReuse(CHI2014).pdf}

\hyperdef{}{ref-Jackson:2013a}{\label{ref-Jackson:2013a}}
42. Steven Jackson, Stephanie Steinhardt, and Ayse Buyuktur. 2013. Why
CSCW needs science policy (and vice versa). In \emph{Proceedings of the
2013 conference on computer supported cooperative work}.

\hyperdef{}{ref-Jasanoff:2006}{\label{ref-Jasanoff:2006}}
43. Sheila Jasanoff. 2006. \emph{States of knowledge: The co-production
of science and the social order}. Routledge.

\hyperdef{}{ref-Jenkinson:1922}{\label{ref-Jenkinson:1922}}
44. Hilary Jenkinson. 1922. \emph{A manual of archive administration
including the problems of war archives and archive making}. Clarendon
Press.

\hyperdef{}{ref-Jordan:1996}{\label{ref-Jordan:1996}}
45. Brigitte Jordan. 1996. Ethnographic workplace studies and CSCW. In
\emph{Human factors in information technology}. Elsevier, 17--42.

\hyperdef{}{ref-Kahle:2007}{\label{ref-Kahle:2007}}
46. Brewster Kahle. 2007. Universal access to all knowledge. \emph{The
American Archivist} 70, 1: 23--31.

\hyperdef{}{ref-Karasti:2010}{\label{ref-Karasti:2010}}
47. Helena Karasti, Karen S. Baker, and Florence Millerand. 2010.
Infrastructure time: Long-term matters in collaborative development.
\emph{Computer Supported Cooperative Work} 19: 377--415.

\hyperdef{}{ref-Kitchin:2016}{\label{ref-Kitchin:2016}}
48. Rob Kitchin. 2016. Thinking critically about and researching
algorithms. \emph{Information, Communication \& Society}: 1--16.

\hyperdef{}{ref-Kuutti:2014}{\label{ref-Kuutti:2014}}
49. Kari Kuutti and Liam J Bannon. 2014. The turn to practice in HCI:
Towards a research agenda. In \emph{Proceedings of the 32nd annual ACM
Conference on Human Factors in Computing Systems}, 3543--3552. Retrieved
from \url{http://dl.acm.org/citation.cfm?id=2557111}

\hyperdef{}{ref-Lasfargues:2012}{\label{ref-Lasfargues:2012}}
50. France Lasfargues, Leila Medjkoune, and Chloé Martin. 2012.
Archiving before loosing valuable data? Development of web archiving in
europe. \emph{Bibliothek Forschung und Praxis} 36, 1: 117--124.

\hyperdef{}{ref-Leetaru:2015a}{\label{ref-Leetaru:2015a}}
51. Kalev Leetaru. 2015. Why it's so important to understand what's in
our web archives. \emph{Why It's So Important To Understand What's In
Our Web Archives}. Retrieved from
\url{http://www.forbes.com/sites/kalevleetaru/2015/11/25/why-its-so-important-to-understand-whats-in-our-web-archives/}

\hyperdef{}{ref-Leetaru:2015}{\label{ref-Leetaru:2015}}
52. Kalev Leetaru. 2015. How much of the internet does the Wayback
Machine really archive? \emph{Forbes}. Retrieved from
\url{http://www.forbes.com/sites/kalevleetaru/2015/11/16/how-much-of-the-internet-does-the-wayback-machine-really-archive/}

\hyperdef{}{ref-Lyons:2016}{\label{ref-Lyons:2016}}
53. Bertram Lyons. 2016. There will be no digital dark age. Retrieved
from
\url{https://issuesandadvocacy.wordpress.com/2016/05/11/there-will-be-no-digital-dark-age/}

\hyperdef{}{ref-Madhavan:2008}{\label{ref-Madhavan:2008}}
54. Jayant Madhavan, David Ko, Łucja Kot, Vignesh Ganapathy, Alex
Rasmussen, and Alon Halevy. 2008. Google's deep web crawl.
\emph{Proceedings of the VLDB Endowment} 1, 2: 1241--1252.

\hyperdef{}{ref-Marill:2004}{\label{ref-Marill:2004}}
55. Jennifer Marill, Andy Boyko, and Michael Ashenfelder. 2004.
\emph{Web harvesting survey}. International Internet Preservation
Consortium. Retrieved from
\url{http://www.netpreserve.org/sites/default/files/resources/WebArchivingSurvey.pdf}

\hyperdef{}{ref-Mohr:2004}{\label{ref-Mohr:2004}}
56. Gordon Mohr, Michael Stack, Igor Rnitovic, Dan Avery, and Michele
Kimpton. 2004. Introduction to heritrix. In \emph{4th international web
archiving workshop}. Retrieved from
\url{https://webarchive.jira.com/wiki/download/attachments/5441/Mohr-et-al-2004.pdf}

\hyperdef{}{ref-Monteiro:2013}{\label{ref-Monteiro:2013}}
57. Eric Monteiro, Neil Pollock, Ole Hanseth, and Robin Williams. 2013.
From artefacts to infrastructures. \emph{Computer Supported Cooperative
Work} 22.

\hyperdef{}{ref-NDSA:2012}{\label{ref-NDSA:2012}}
58. NDSA. 2012. \emph{Web archiving survey report}. National Digital
Stewardship Alliance. Retrieved from
\url{http://www.digitalpreservation.gov/ndsa/working_groups/documents/ndsa_web_archiving_survey_report_2012.pdf}

\hyperdef{}{ref-Neubert:2014}{\label{ref-Neubert:2014}}
59. Michael Neubert. 2014. Five questions for will elsbury, project
leader for the election 2014 web archive. Retrievedfrom
\url{http://blogs.loc.gov/digitalpreservation/2014/10/five-questions-for-will-elsbury-project-leader-for-the-election-2014-web-archive/}

\hyperdef{}{ref-Niu:2012b}{\label{ref-Niu:2012b}}
60. Jinfang Niu. 2012. Appraisal and custody of electronic records:
Findings from four national archives. \emph{Archival Issues} 34, 2:
117--130.

\hyperdef{}{ref-Niu:2012a}{\label{ref-Niu:2012a}}
61. Jinfang Niu. 2012. An overview of web archiving. \emph{D-Lib
magazine} 18, 3. Retrieved from
\url{http://www.dlib.org/dlib/march12/niu/03niu1.html}

\hyperdef{}{ref-Pendse:2014}{\label{ref-Pendse:2014}}
62. Liladhar R Pendse. 2014. Archiving the Russian and East European
lesbian, gay, bisexual, and transgender web, 2013: A pilot project.
\emph{Slavic \& East European Information Resources} 15, 3: 182--196.

\hyperdef{}{ref-Phillips:2006}{\label{ref-Phillips:2006}}
63. Margaret E Phillips. 2006. What should we preserve? The question for
heritage libraries in a digital world. \emph{Library trends} 54, 1:
57--71.

\hyperdef{}{ref-Ribes:2009}{\label{ref-Ribes:2009}}
64. David Ribes and Thomas Finholt. 2009. The long now of technology
infrastructure: Articulating tensions in development. \emph{Journal of
the Association for Information Systems} 10, 5.

\hyperdef{}{ref-Ribes:2010}{\label{ref-Ribes:2010}}
65. David Ribes and Charlotte P. Lee. 2010. Sociotechnical studies of
cyberinfrastructure and e-research: Current themes and future
trajectories. \emph{Computer Supported Cooperative Work (CSCW)} 19, 3:
231--244.
\href{http://doi.org/10.1007/s10606-010-9120-0}{https://doi.org/10.1007/s10606-010-9120-0}

\hyperdef{}{ref-Rosner:2014}{\label{ref-Rosner:2014}}
66. Daniela K Rosner and Morgan Ames. 2014. Designing for repair?:
Infrastructures and materialities of breakdown. In \emph{Proceedings of
the 17th ACM conference on computer supported cooperative work \& social
computing}, 319--331. Retrieved from
\url{http://people.ischool.berkeley.edu/~daniela/files/cscw14-rosner-repair.pdf}

\hyperdef{}{ref-Saad:2012}{\label{ref-Saad:2012}}
67. Myriam Ben Saad and Stéphane Gançarski. 2012. Archiving the web
using page changes patterns: A case study. \emph{International Journal
on Digital Libraries} 13, 1: 33--49.

\hyperdef{}{ref-Samuels:1986}{\label{ref-Samuels:1986}}
68. Helen Willa Samuels. 1986. Who controls the past. \emph{The American
Archivist}: 109--124. Retrieved from
\url{http://americanarchivist.org/doi/abs/10.17723/aarc.49.2.t76m2130txw40746}

\hyperdef{}{ref-Sanderson:2011}{\label{ref-Sanderson:2011}}
69. Robert Sanderson, Mark Phillips, and Herbert Van de Sompel. 2011.
Analyzing the persistence of referenced web resources with Memento.
Retrieved from \url{http://arxiv.org/abs/1105.3459}

\hyperdef{}{ref-Schellenberg:1956}{\label{ref-Schellenberg:1956}}
70. Theodore R Schellenberg. 1956. \emph{Modern archives: Principles and
techniques}. University of Chicago Press. Retrieved from
\url{http://catalog.hathitrust.org/Record/003147122}

\hyperdef{}{ref-Seaver:2013}{\label{ref-Seaver:2013}}
71. Nick Seaver. 2013. Knowing algorithms. \emph{Media in Transition} 8:
1--12. Retrieved from \url{http://nickseaver.net/s/seaverMiT8.pdf}

\hyperdef{}{ref-Star:1999}{\label{ref-Star:1999}}
72. Susan Leigh Star. 1999. The ethnography of infrastructure.
\emph{American behavioral scientist} 43, 3: 377--391.

\hyperdef{}{ref-Suchman:1986}{\label{ref-Suchman:1986}}
73. Lucy Suchman. 1985. \emph{Plans and situated actions: The problem of
human-machine communication.} Xerox Corporation.

\hyperdef{}{ref-Szydlowski:2010}{\label{ref-Szydlowski:2010}}
74. Nick Szydlowski. 2010. Archiving the web: It's going to have to be a
group effort. \emph{The Serials Librarian} 59, 1: 35--39.

\hyperdef{}{ref-Taylor:1992}{\label{ref-Taylor:1992}}
75. Hugh A. Taylor. 1992. \emph{The archival imagination: Essays in
honour of Hugh A. Taylor}. Association of Canadian Archivists.

\hyperdef{}{ref-Trace:2010}{\label{ref-Trace:2010}}
76. Ciaran B Trace. 2010. On or off the record? Notions of value in the
archive. \emph{Currents of Archival Thinking}: 47--68.

\hyperdef{}{ref-Tschan:2002}{\label{ref-Tschan:2002}}
77. Reto Tschan. 2002. A comparison of jenkinson and schellenberg on
appraisal. \emph{The American Archivist} 65, 2: 176--195.

\hyperdef{}{ref-Yaco:2015}{\label{ref-Yaco:2015}}
78. Sonia Yaco, Ann Jimerson, Laura Caldwell Anderson, and Chanda
Temple. 2015. A web-based community-building archives project: A case
study of kids in birmingham 1963. \emph{Archival Science} 15, 4:
399--427.

\hyperdef{}{ref-Zittrain:2014}{\label{ref-Zittrain:2014}}
79. Jonathan Zittrain, Kendra Albert, and Lawrence Lessig. 2014. Perma:
Scoping and addressing the problem of link and reference rot in legal
citations. \emph{Legal Information Management} 14, 02: 88--99.

\end{document}